\documentclass[a4paper,18pt]{report}
\usepackage{graphicx}
\usepackage[usenames]{color}
\newcommand{\HoTi}{Ho$_2$Ti$_2$O$_7$ }
\newcommand{\DyTi}{Dy$_2$Ti$_2$O$_7$ }
\newcommand{\DyTins}{Dy$_2$Ti$_2$O$_7$}
\newcommand{\HoTins}{Ho$_2$Ti$_2$O$_7$}
\newcommand\degrees{\ensuremath{^\circ}}

\begin{document}

{\noindent{\large{\bf Pinch Points and Kasteleyn Transitions: How
Spin Ice Changes its Entropy}}} \vskip 0.5cm

{\noindent T. Fennell$^{1,2}$, S. T. Bramwell$^{1,3}$, D. F.
McMorrow$^{1,2,4}$ P. Manuel$^{4}$}

\vskip 0.5cm {\noindent $^{1}$London Centre for Nanotechnology,
2-16 Torrington Place, London WC1E 7HN, UK; \quad $^{2}$Department
of Physics, University College London, Gower Street, London, WC1E
6BT, UK; \quad $^{3}$Department of Chemistry, University College
London, 20 Gordon Street, London WC1H 0AJ, UK; \quad $^{4}$ISIS
Facility,
Rutherford Appleton Laboratory, Chilton, Didcot, Oxon, OX11 0QX, UK}\\

\begin{figure}
\begin{minipage}[c]{0.5\textwidth}
    \includegraphics[scale=0.5]{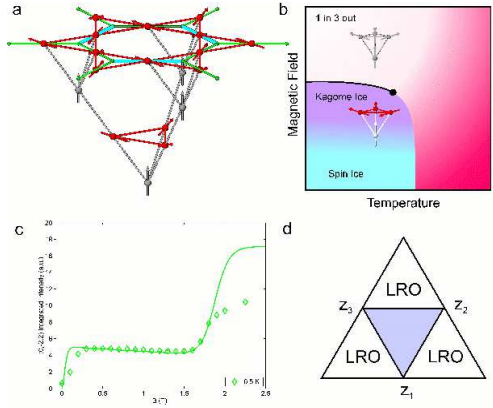}
\end{minipage}
\caption{\label{phasemap} {\small Spin ice and kagome ice. With
the trigonal axis ([111]) vertical, the pyrochlore lattice can be
viewed as stacked kagome planes, separated by interstitial spins
forming a triangular lattice ({\bf a}).  In zero field all
tetrahedra have a sixfold degenerate groundstate with two spins
pointing in and two pointing out (i.e. ice rules), in moderate
field this degeneracy is reduced by the pinning of one spin
leaving a modified ice rule operating on the three spins in the
kagome plane ({\bf a} and {\bf b}, lower tetrahedron), in high
fields this ice rule is broken and a fully ordered structure with
one spin pointing in and three spins pointing out ({\bf b}, upper
tetrahedron) is formed.  The transition from kagome ice to long
range order shows a liquid-gas critical point in \DyTi [5] and in
\HoTi a critical point of unknown type (see main text and
Fig.~\ref{tp}).  The three phases are distinguished in the
magnetic Bragg scattering ({\bf c}) which shows two plateaux, the
first for kagome ice and the second for the fully ordered state.
The line is a calculation of the intensity by enumeration of
states on a single tetrahedron with field close to [111], as
described in the text.  The tilted field is responsible for the
downward trending intensity with field within the kagome ice
plateau. Lack of agreement at high fields is attributed to severe
extinction at this very intense magnetic Bragg peak.  The kagome
ice phase can be mapped to the dimer model on a hexagonal lattice
by colouring links of the hexagonal lattice located on sites of
the kagome lattice with field-opposing spins ({\bf a}, hexagonal
dual lattice in green, dimers in cyan) [23]. The dimer model has a
triangular phase diagram with respect to statistical weights of
different dimer orientations, the central region corresponds to
kagome ice and the lines correspond to Kasteleyn transitions to
long range ordered states (LRO) which occur when one dimer
orientation outweighs the other two ({\bf d}).}}
\end{figure}

{\noindent{\bf Complex disordered states - from liquids and
glasses to exotic quantum matter - are ubiquitous in nature. Their
key properties include finite entropy, power-law correlations and
emergent organising principles. In spin ice, spin correlations are
determined by an ice rules organising principle that stabilises a
magnetic state with the same zero point entropy as water ice. The
entropy can be manipulated with great precision by a magnetic
field: with field parallel to the trigonal axis one obtains quasi
two dimensional kagome ice which can be mapped onto a dimer model.
Here we use a field tilted slightly away from the trigonal axis to
control the dimer statistical weights and realise the unusual
critical behaviour predicted by Kasteleyn. Neutron scattering on
Ho$_2$Ti$_2$O$_7$ reveals pinch point scattering that
characterises the emergent gauge structure of kagome ice; diffuse
peaks that shift with field, signaling the Kasteleyn physics; and
an unusual critical point.  }}
\newpage
In spin ice materials such as \HoTi and \DyTi [1-6], the
Ising-like rare earth spins are analogous to hydrogen displacement
vectors in water ice, so the statistical mechanics of these
materials may be mapped directly onto Pauling's model of hydrogen
disorder in water ice. The ground state organising principle, or
``ice rule'' is that two spins should point into and two out of
each elementary tetrahedron of the lattice occupied by Ho or Dy
(Fig.~\ref{phasemap}a and b). This rule does not enforce
long-range spin order and spin ice shares with water ice the
Pauling zero point entropy [2,7].  A magnetic field directed along
the trigonal axis (Fig.~\ref{phasemap}b) ``pins'' one spin per
tetrahedron along the field direction, but, provided that it is
not too strong, allows the ice rules to dictate that the other
three spins per tetrahedron must be organised into
``two-out-one-in'' (for an ``up'' tetrahedron as pictured). As
first observed by Matsuhira et al. [4], this modified ice rule
results in a reduced, but still well-defined, zero point entropy.
The spins carrying this entropy are arranged in decoupled
two-dimensional sheets with kagome geometry, hence the name
``kagome ice'' [4]. Application of a stronger field breaks the ice
rule to enforce an ordered ``one-in-three-out'' arrangement,
giving the magnetization a two plateau form (Fig.~\ref{phasemap}b
and c). This process is accompanied by a giant spike in the
residual entropy [8,9].  The entropy spike is due to
the crossing of an extensive number of levels which have
macroscopic entropies: at the critical field all possible kagome
ice states, one-in-three-out states and combinations of the two,
have the same energy as the field exactly balances the
interactions [7]. This process, predicted by Bonner and Fisher
[10], is anticipated to be general for any Ising antiferromagnet
but remains largely uninvestigated experimentally.

The properties of the spin ice materials have attracted much
theoretical interest. In spin ice  the ice rules have been 
shown to arise
as a many body effect of the dipolar interaction between magnetic
moments [11-14].  What is remarkable about spin ice is that the
ice rules emerge from the dipolar interaction, but in other cases
an effective dipolar interaction may emerge from ice rules
stabilised by local interactions as in the case of water ice or
the near neighbour antiferromagnet [7,15]. In either situation the
key experimental signatures are ``pinch point'' singularities in
the static scattering function $S(\mathbf{Q})$ [13,16,17]. Because
ice rules are a ``Gauss's law'' or divergence free local
constraint, they can generally stabilise spin patterns that are
analogous to magnetic or electric field interactions, which
means that longitudinal magnetization fluctuations are completely
suppressed: a so-called ``emergent gauge structure''. This
produces the pinch point scattering, which is singular in one
direction and diffuse in all others. Well defined pinch points
have previously been observed in ferroelectric systems like KDP
[18] where $S(\mathbf{Q})$ is a measure of the local polarization
fluctuations.  However, they are less well defined in candidate
magnetic systems like CsNiCrF$_6$ [19], (Y$_{1-x}$Sc$_x$)Mn$_2$
[20] and spin ice in zero field [21,22]. Our observation of them
(described below) may be the first clear example in magnetism.

Moessner and Sondhi showed that the  kagome ice phase can be
mapped to the dimer model on the hexagonal lattice [23].
Every triangle of the kagome lattice has two
spins with positive projection on the field and one opposed to the
field.  If the kagome lattice is replaced by its hexagonal dual
and the bonds centered on a field-opposing spin are coloured, a
kagome ice state becomes a disordered dimer state on the hexagonal
lattice, as illustrated in Fig.~\ref{phasemap}a. The dimer model
on the hexagonal lattice was orginally studied by Kasteleyn [24],
who showed that the dimer correlations are critical. Three dimer
orientations are available and Kasteleyn found a triangular phase
diagram depending on their statistical weights $Z_1$, $Z_2$ and
$Z_3$ (Fig.~\ref{phasemap}d). When $Z_1>Z_2=Z_3$ (for example),
the first dimer orientation will be selected and there will be a
transition to a long range ordered dimer solid.  This
so-called Kasteleyn transition has many remarkable properties,
elucidated by Moessner and Sondhi [23], including an asymmetric
first/second order appearance (as yet unobserved in experiment).
In kagome ice it is expected that small tilts of the field close
to the [111] axis will be equivalent to tuning the statistcal
weights of the dimer orientations [23].

\begin{figure}
\begin{minipage}[c]{0.5\textwidth}
    \includegraphics[scale=0.75]{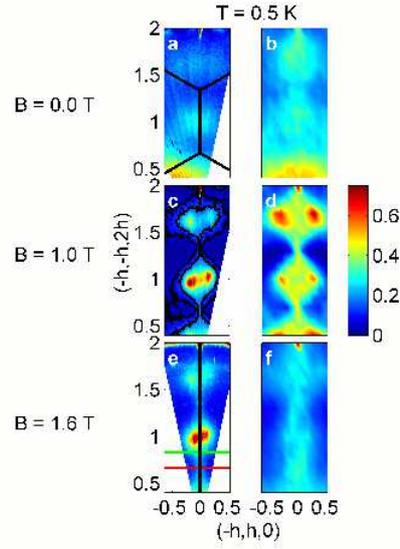}
\end{minipage}
\caption{\label{datamaps} Measured and simulated diffuse
scattering in \HoTins.  In zero field the experimental diffuse
scattering ({\bf a}, 0.0 T, 0.5 K) is only approximately matched
by the nearest neighbour spin ice model ({\bf b}), it is
anticipated that a dipolar spin ice simulation would be a
significant improvement [11,21,26]. In the kagome ice phase ({\bf
c}, 1.0 T, 0.5 K) very clear pinch points are observed at
$(\bar{\frac{2}{3}},\bar{\frac{2}{3}},\frac{4}{3})$ and
$(\bar{\frac{4}{3}},\bar{\frac{4}{3}},\frac{8}{3})$.  Remarkably
the agreement with the nearest neighbour model ({\bf d}) is much
improved. The splitting of the central peak is less than that
predicted by the simulation and this is ascribed to the tilt. At
the termination of the plateau the scattering is localized in a
single sharp feature at $(\bar{1},\bar{1},2)$ ({\bf e}, 1.6 T, 0.5
K) which has the features of an unusual critical point.  This
behaviour is not captured by the nearest neighbour model ({\bf
d}). }
\end{figure}

\begin{figure}
\begin{minipage}[c]{0.5\textwidth}
    \includegraphics[scale=0.75]{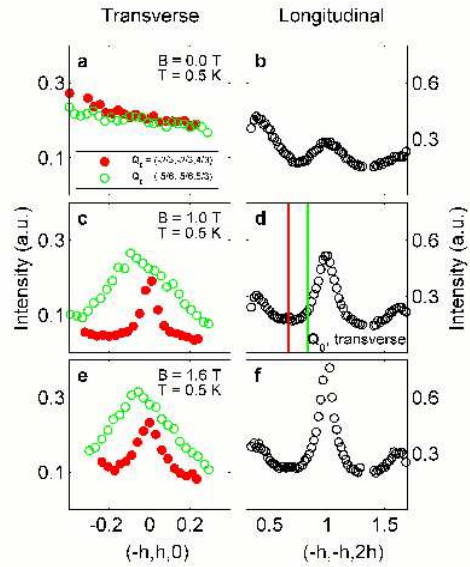}
\end{minipage}
\caption{\label{cuts} Pinch point scattering.  Cuts through the
pinch point (the cut positions are illustrated in matching colours
in Fig.~\ref{datamaps}e, all errorbars are smaller than the symbol
size) position in zero field ({\bf a},{\bf b}) show spin ice
scattering and no sharp feature in the transverse cuts parallel to
$(\bar{h},h,0)$ ({\bf a}). In 1 T the sharp pinch point has been
formed ({\bf c},{\bf d}) and also exists in 1.6 T ({\bf e},{\bf
f}).  The pinch point can be seen as the sharp feature in the cut
at $(\bar{h},h,\frac{4}{3})$, comparatively broadened at
$(\bar{h},h,\frac{9}{6})$ and diffuse in the cut along
$(\bar{h},\bar{h},2h)$ where the pinch point position is
$(\bar{\frac{2}{3}},\bar{\frac{2}{3}},\frac{4}{3})$. In the
longitudinal cuts ({\bf b},{\bf d},{\bf f}) we also see the
developing critical scattering at $(\bar{1},\bar{1},2$).}
\end{figure}

We performed neutron scattering experiments on holmium titanate,
\HoTins. Although detailed bulk measurements of the $[111]$ field
direction have mainly been performed on dysprosium titanate,
\DyTins, the holmium material lends itself much better to neutron
scattering and it is expected to show qualitatively the same
behaviour.  In general, the diffuse neutron scattering is a direct
measure of $S(\mathbf{Q})$, the Fourier transformed spin-spin
correlation function: it would show resolution limited Bragg peaks
for long range order, broad diffuse peaks for short range order
and sharp, but non-resolution limited peaks for critical
scattering. Fig.~\ref{phasemap}c shows the evolution of the
$(0,\bar{2},2)$ magnetic Bragg peak intensity of \HoTi as a
function of field. This quantity shows a similar two plateau
structure to the bulk magnetization [25] with the lower plateau
corresponding to the kagome ice phase. Fig.~\ref{datamaps}a, c and
e show the corresponding neutron scattering patterns in the plane
of reciprocal space perpendicular to $[111]$ in zero field, in the
kagome ice plateau and at the termination of the plateau
respectively. In the kagome ice phase the diffuse scattering
compares quite well with a simulation of the near neighbour spin
ice model (Fig.~\ref{datamaps}c), with two broad peaks and a
saddle point at to $(\bar{1},\bar{1},2)$ and striking pinch point
singularities at
$(\bar{\frac{2}{3}},\bar{\frac{2}{3}},\frac{4}{3})$ and
$(\bar{\frac{4}{3}},\bar{\frac{4}{3}},\frac{8}{3})$.  The position
of the pinch points corresponds to the zone centers of a
$\mathbf{Q}=0$ cell in the kagome plane. Proof that these are
indeed pinch points in the manner discussed in Refs. 13, 16, and
17 is given in Fig.~\ref{cuts}. Such pinch points are a direct
signature of a state governed by ice-rules or an emergent gauge
structure.  The remarkable feature is the ease with which these
pinch points can be detected. They are explicitly expected to
appear in the $S(\mathbf{Q})$ of spin ice [13,14] but have not
been observed in experiment [21,22], leading to suggestions that
they are hidden by other features or too weak to detect at finite
temperature. Here the pinch points are extremely clear, posing the
question of how ice rule governed spin correlations are modified
on dimensional reduction.  Very recently Tabata {\it et al} [26]
have published a neutron scattering study of \DyTi with field
parallel to [111] and apparently zero tilt. Within the constraint
of the strong neutron absorption by Dy they are able to confirm
the existence of the kagome ice phase, but their data lacks the
resolution to test for pinch point singularities or the other
physics revealed in our study of \HoTins.

We next discuss the Kasteleyn physics.  In order to
observe the Kasteleyn transition as described by Moessner and
Sondhi [23], one must tilt the field toward $[112]$.  Considering
an ``up'' tetrahedron, with [111] vertical, this corresponds to
increasing the effective field at two sites.  The remaining site
will preferentially carry the ``in'' spin or dimer.  Tilting the
field therefore corresponds to tuning the statistical weights of
dimer orientations in Kasteleyn's model, as described above (i.e.
$Z_1>Z_2=Z_3$).  In this situation it is expected that as a
function of tilt, the broad peaks would drift away from their
positions toward the $(0,\bar{2},2)$ and $(\bar{2},0,2)$
positions, reaching the zone centers at the transition.  On the
Kasteleyn phase diagram (Fig.~\ref{phasemap}d), tilt toward
$[112]$ crosses a phase boundary into a dimer solid phase. The
selection of a single dimer orientation simultaneously on every
stacked kagome layer must therefore generate long range order, in
this case equivalent to the $Q=0$ order observed with the field
parallel to $[001]$, in which magnetic Bragg scattering is
observed only at the zone centers [1,27].

If the tilt is toward $[110]$, dimer weights are still altered. In
this case the single spin selected will preferentially maintain a
positive projection on the field direction.  This is equivalent to
excluding dimer occupation from this site, or $Z_1<Z_2=Z_3$. On
the Kasteleyn phase diagram, tilting toward [110] crosses the
point at which two dimer solids meet. The dimer phase formed is
partially ordered with uncorrelated chains of either of the two
possible dimer orientations. In spin language, this corresponds to
the $\mathbf{Q}=X$ partial order previously studied in both \HoTi
and \DyTi [27]. The tilted field creates an in-out pair from two
spins which form a chain, the ice rules form in-out chains from
the remaining spins.

\begin{figure}
\begin{minipage}[c]{0.5\textwidth}
    \includegraphics[scale=0.75]{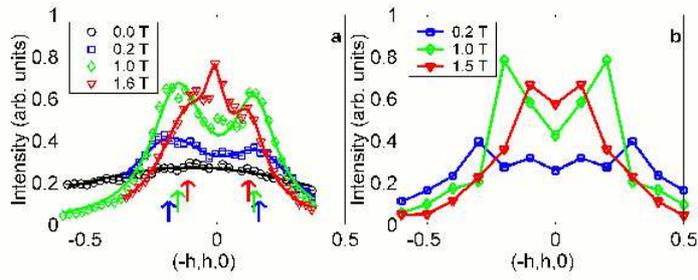}
\end{minipage}
\caption{\label{mccuts} Kasteleyn physics. Cuts through the main
scattering feature at $(\bar{1},\bar{1},2)$ as a function of field
show that the broad feature of zero field is replaced by two peaks
which drift together as the field is raised across the kagome ice
plateau.  The experimental data show this process, with the
apparent superimposition of the sharp critical like scattering at
1.6 T ({\bf a}), whereas Monte Carlo simulations of the same field
scan with a small tilt, show only the drifting peaks as the
critical scattering is not reproduced by the nearest neighbour
model ({\bf b}).  The fitted peak positions in the experimental
data are marked below by the arrows.}
\end{figure}

The tuning of dimer weights goes approximately as $\phi B/T$
($\phi$ is the tilt angle) so from a position close to the center
of the critical phase, one can move toward the phase boundaries by
increasing field or tilt, or lowering temperature. Our study is at
fixed tilt of about $1\degrees$ toward $[110]$ and we vary the
field (which is applied approximately parallel to
$[1.05\quad1.03\quad1]$).  In Fig.~\ref{mccuts} we compare the
scattering at 0.2, 0.5, 1.0 and 1.6 T and see that the split peaks
sharpen and move inward, a clear signature of the Kasteleyn tuning
of the critical phase.  This is also manifested in the Bragg
intensity (Fig.~\ref{phasemap}c), which decreases slightly across
the plateau as a function of $B$.  This is due to the preferential
location of the ``in'' spin at a particular site.  Monte Carlo
simulations of the nearest neighbour spin ice model with varying
tilt toward $[110]$, confirm that the peaks drift inward toward
the $(\bar{1},\bar{1},2)$ when the tilt is in this sense
(Fig.~\ref{mccuts}b). In the nearest neighbour model the chains
are not coupled, hence only one dimensional order is achieved
a rod of scattering appears, but it is known in the real systems 
that
highly anisotropic partial order appears, with non-resolution
limited features appearing at $\mathbf{Q}=X$ positions such as
$(\bar{1},\bar{1},2)$ [27].
 Although we have observed a key signature of Kasteleyn's predictions,
we have not explicitly observed the transition. However, the
coincident long range or partial order according to tilt
direction, and their manifestation at different wavevectors
promises a very clear method by which this can be
achieved.

\begin{figure}
\begin{minipage}[c]{0.5\textwidth}
\includegraphics[scale=0.75]{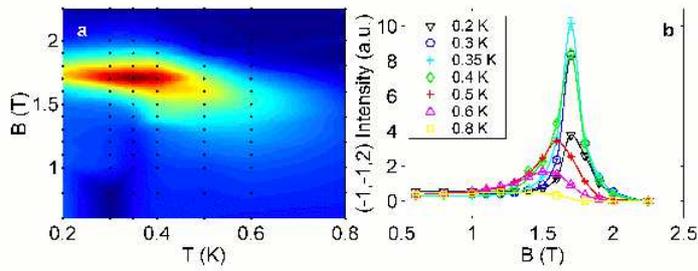}
\end{minipage}
\caption{\label{tp} Critical point at plateau termination.  The
intensity of the critical scattering at $\bar{1},\bar{1},2$ is
strongly dependent on temperature, having a strong maximum at $T =
0.35$ K and $B = 1.7$ T, giving the appearance of a critical
point.  We show the phase boundary in {\bf a} (note the intensity
scale is logarithmic) and the detailed behaviour of the intensity
in {\bf b}. It is notable that below $T_c$ the intensity is field
independent in the plateau until close to $B_c$ whereas above
$T_c$ the intensity increases throughout the plateau toward its
maximum at $B_c$.}
\end{figure}

An applied field of 1.6 T corresponds to the transition region
between the two magnetization plateaus where the
``two-in-two-out'' rule is broken and one expects a giant entropy
spike [7,8]. In \DyTi [6], though not in the near neighbour model,
this transition takes the form of a line of first order phase
transitions terminating in a critical end point on the field $(B)$
versus temperature $(T)$ phase diagram (Fig.~\ref{phasemap}b). The
experimental entropy peaks along this line and reaches a maximum
at the end point. We find, for \HoTins, an anomalous growth in
neutron scattering at the point $(\bar{1},\bar{1},2)$ that peaks
near to the expected first order line where it completely swamps
the remnants of the kagome ice scattering (see Figs.~\ref{cuts}
and~\ref{tp}). This scattering is relatively sharp in reciprocal
space although never a Bragg peak: it closely resembles critical
scattering.  It possibly indicates the chain state discussed
above, but does not seem to depend on tilt, associating it more
probably with the transition to the ``one-in-three-out'' state
that terminates the plateau.  Indeed, it behaves much like the
expected entropy, having a strong peak at a single $B,T$ point
(Fig.~\ref{tp}). These features strongly suggest a critical point
at $B\approx 1.6$ T, $T = 0.35$ K, $\mathbf{Q} =
(\bar{1},\bar{1},2)$.  Above $T_c=0.35$ K the intensity at
$\bar{1},\bar{1},2$ is continuously growing as the field is
scanned across the plateau, but below $T_c$ the intensity in the
plateau is independent of field until close to 1.6 T, when there
is a sharp onset of the transition.

This critical point is unusual. At a normal critical point, one
would expect critical scattering to precede the development of
long range order with a Bragg peak at the same, ordering
wavevector. In the expected symmetry sustaining transition,
critical scattering near to the $\mathbf{Q} = 0$ positions should
accompany changes in Bragg peak intensity at these positions.
However, we see strong critical scattering only at the
antiferromagnetic wavevector $(\bar{1},\bar{1},2)$ and in no part
of the $(B,T)$ phase diagram do we see a Bragg peak develop at
this position. Why an antiferromagnetic ordering wavevector should
go critical, remains, for the time being, a mystery, but the
correlation of the $(\bar{1},\bar{1},2)$ peak with the expected
entropy suggests an association of these two properties. Entropy
peaks are a general feature of Ising model systems and result from
extra degeneracies that are introduced when the field just
balances local interactions [9,8,23]. The extra degeneracy
corresponds to extra spin configurations added to the manifold of
thermally accessible states: here it appears that these
configurations are locally antiferromagnetic.

To summarize our results, we have observed the long sought pinch
points, characteristic of an ice rules governed magnet. These lie
at the zone centers of a $Q=0$ cell in the kagome plane. It is a
fascinating challenge for theorists to resolve their apparent
absence in three dimensional spin ice, with their presence in two
dimensional kagome ice.  We have observed the first experimental
signature of Kasteleyn's tunable dimer critical phase, uncovered a
new type of dimer ordering that can be realized by this mechanism,
and demonstrated an open route to the realization of the full
Kasteleyn transition. Finally we have observed an unusual critical
point, at which the critical wavevector does not reflect the
developing long range order.

In conclusion, the concept of spin ice is relevant to a wide
variety of highly correlated systems, including frustrated magnets
[16,28,29,30], disordered magnets [31], anomalous metals [32],
nanomagnetic arrays [33] and ice itself (which has been described
as a highly correlated proton system) [34].  With a field applied
along [111], spin ice exhibits some generic phenomena. For
example, first order lines with critical end points play an
important role in metal-insulator transitions [35] and
ferroelectric response [36], while partially ordered phases on the
kagome lattice are also predicted for systems of bosons [37]. Our
neutron scattering investigation of holmium titanate's kagome ice
plateau has thus revealed a detailed microscopic picture of the
statistical physics of a model system that has broad relevance to
many aspects of the physics of complex disorder.

\vskip 1.5cm

{\bf Methods:} A flux-grown single crystal of \HoTi of dimensions
$3\times10\times15$ mm was aligned with the $[111]$ axis vertical.
In this orientation the scattering plane has threefold symmetry
and is spanned by the orthogonal sets of vectors
$(\bar{h},\bar{h},2h)$ and $(\bar{h},h,0)$.  The crystal was
mounted on a dilution refrigerator insert in a 7 T vertical field
cryomagnet. The PRISMA spectrometer at the ISIS facility was used
in diffraction mode to map the scattering plane or to make rocking
scans about the high symmetry directions (as in [20]).
Optimization of the scattering plane while monitoring the
intensity of $(\bar{2},\bar{2},4$ and $(2,\bar{2},0)$ shows that
these axes are tilted $2\degrees$ above and $1\degrees$ below the
scattering plane respectively. This corresponds to an applied
field approximately parallel to $[1.05\quad1.03\quad1]$, a small
tilt toward the $[110]$.  The near neighbour spin ice model was
simulated using a standard Metropolis algorithm.  The system size
was $10\times10\times10$ pyrochlore unit cells.  The system was
cooled stepwise to low temperature to ensure equilibration in the
spin ice manifold, 40000 Monte Carlo steps per spin were used at
each temperature for equilibration, followed by 200000 Monte Carlo
steps per spin for accumulating thermodynamic quantities, which
were sampled every 10000 steps.  A series of
 five such simulations with different seed configurations were averaged. The
tilted fields were introduced as
$\mathbf{B}=B(\cos\phi[111]/\sqrt3+\sin\phi[\bar{1}\bar{1}2]/\sqrt6)$,
either at fixed $\phi$ throughout, or by stepping $\phi$ at fixed
B.

\vskip 0.5cm {\bf Acknowledgements:} It is a pleasure to
acknowledge sample environment team at ISIS (in particular R. Down
and J. Keeping) and we
would like to thank A.S. Wills and R. Moessner for valuable
discussions. We thank the EPSRC (UK) for funding.

\newpage
{\noindent{\bf References}}

[1] Harris, M. J., Bramwell, S. T., McMorrow, D. F., Zeiske, T.,
Godfrey, K. W., Geometric Frustration in the Ferromagnetic
Pyrochlore \HoTins,~{\it Phys. Rev. Lett.} {\bf 79}, 2554 (1997).

[2] Ramirez, A. P., Hayashi A., Cava, R. J., Siddharthan R. \&
Shastry, B. S. Zero-point entropy in `spin ice'. {\it Nature} {\bf
399,} 333 (1999).

[3] Snyder, J., Slusky, J. S., Cava, R. J. \& Schiffer P. How
`spin ice' freezes. {\it Nature} {\bf 413,} 48 (2001).

[4] Matsuhira, K., Hiroi,  Z., Tayama,  T., Takagi, S. \&
Sakakibara,  T. A new macroscopically degenerate ground state in
the spin ice compound \DyTins. {\it  J. Phys. Condens. Matter}
{\bf 14,} L559 (2002).

[5] Sakakibara, T., Tayama, T., Hiroi, Z., Matsuhira, K. \& Takagi
S., Observation os Liquid-Gas-Type Transition if the Pyrochlore
Spin Ice Compound \DyTi in a Magnetic Field. {\it Phys. Rev.
Lett.} {\bf 90,} 207205 (2003).

[6] Higashinaka R. \& Maeno, Y. Field-Induced Transition of a
Triangular Plane in the Spin-Ice Compound \DyTins. {\it Phys. Rev.
Lett.} {\bf 95,} 237208 (2005)

[7] Pauling, L., The Structure and Entropy of Ice and of Other
Crystals with Some Randomness of Atomic Arrangement. {\it J. Am.
Chem. Soc.} {\bf 57,} 2680 (1935).

[8] Isakov, S. V., Raman,  K. S. , Moessner, R. \&    Sondhi, S.
L. Magnetization curve of spin ice in a [111] magnetic field. {\it
Phys. Rev. B} {\bf 70,} 104418 (2004)

[9] Aoki, H. ,Sakakibara,  T., Matsuhira  K. \& Hiroi, Z.
Magnetocaloric Effect Study on the Pyrochlore Spin Ice Compound
\DyTi in a [111] Magnetic Field. {\it J. Phys. Soc. Japan} {\bf
73,} 2851 (2004).

[10] Bonner, J. C. \& Fisher, M. E. Entropy of an antiferromagnet
in a magnetic field. {\it Proc. Phys. Soc.} {\bf 80,} 508 (1962).

[11] den Hertog, B. C. \& Gingras, M. J. P. Dipolar Interactions
and Origin of Spin Ice in Ising Pyrochlore Magnets. {\it Phys.
Rev. Lett.} {\bf 84,} 3430 (2000).

[12] Melko, R. G. \&  Gingras, M. J. P. Monte Carlo studies of the
dipolar spin ice model. {\it J. Phys. Condens. Matter,} {\bf 16}
R1277-R1319 (2004).

[13] Isakov, S. V., Gregor, K., Moessner, R. \& Sondhi,  S. L.
Dipolar Spin Correlations in Classical Pyrochlore Magnets. {\it
Phys. Rev. Lett.} {\bf 93,} 167204 (2004).

[14] Isakov, S. V. , Moessner, R. \& Sondhi,  S. L. Why Spin Ice
Obeys the Ice Rules. {\it Phys. Rev. Lett.} {\bf 95,} 217201
(2005).

[15] Anderson, P. W., Ordering and Antiferromagnetism in Ferrites.
{\it Phys. Rev.} {\bf 102,} 1008 (1956)

[16] Henley, C. L. Power-law spin correlations in pyrochlore
antiferromagnets. {\it Phys. Rev. B} {\bf 71,} 014424 (2005).

[17] Youngblood, R. W. \& Axe, J. D. Polarization fluctuations in
ferroelectric models {\it Phys. Rev. B} {\bf 23,} 232 (1981).

[18] Skalyo J, Frazer B. C. \& Shirane, G. , Ferroelectric mode
motion in KD$_2$PO$_4$ {\it Phys. Rev. B} {\bf 1,} 278 (1970).

[19] Harris M. J., Zinkin, M. P., Tun, Z., Wanklyn, B. M. \&
Swainson, I. P. Magnetic Structure of the Spin-Liquid state in a
Frustrated Pyrochlore. {\it Phys. Rev. Lett.} {\bf 73} 189 (1994).

[20] Ballou, R., Leli\'evre-Berna, E. \& F{\aa}k, B. Spin
Fluctuations in (Y$_{0.97}$Sc$_{0.03})$Mn$_2$: A Geometrically
Frustrated, Nearly Antiferromagnetic, Itinerant Electron System.
{\it Phys. Rev. Lett.} {\bf 76,} 2125 (1996).

[21] Bramwell, S. T. {\it et al.} Spin correlations in \HoTins: A
Dipolar Spin Ice. {\it Phys. Rev. Lett.,}  {\bf 87,} 047205
(2001).

[22] Fennell, T., {\it et al} Neutron Scattering investigation of
the spin ice state in \DyTins. {\it Phys. Rev. B} {\bf 70,} 134408
(2004).

[23] Moessner, R. \& Sondhi, S. L., Theory of the [111]
magnetization plateau in spin ice. {\it Phys. Rev. B} {\bf 68,}
064411 (2003).

[24] Kasteleyn, P. W. Dimer Statistics and Phase Transitions. {\it
J. Math. Phys.} {\bf 4,} 287 (1963).

[25] Petrenko, O. A., Lees, M. R. \& Balakrishnan, G.
Magnetization process in the spin-ice compound \HoTins. {\it Phys.
Rev. B} {\bf 68,} 012406 (2003).

[26] Tabata, Y. {\it et al.} Kagome ice state in the dipolar spin
ice \DyTins.  arXiv:cond-mat/060708 (2006)

[27] Fennell, T. {\it et al.} Neutron scattering studies of the
spin ices \HoTi and DyTi in applied magnetic field. {\it Phys.
Rev. B} {\bf 72,} 224411 (2005).

[28] Mirebeau, I. {\it et al.} Ordered spin ice state and magnetic
fluctuations in Tb$_2$Sn$_2$O$_7$. {\it Phys. Rev. Lett.} {\bf
94,} 246402 (2005).

[29] Gardner, J. S. {\it et al.} Cooperative paramagnetism in the
geometrically frustrated antiferromanget Tb$_2$Ti$_2$O$_7$.  {\it
Phys. Rev. Lett.} {\bf 82,} 1012 (1999).

[30] Lee, S.-H. {\it et al.} Emergent excitations in a
geometrically frustrated antiferromagnet. {\it Nature} {\bf 418,}
856 (2002).

[31] Lau, G. C., {\it et al} Zero-point entropy in stuffed
spin-ice. {\it Nature Physics} {\bf 2,} 249 (2006).

[32] Taguchi, Y., Oohara, Y., Yoshizawa, H., Nagaosa, N. \& Tokura
Y. Spin Chirality, Berry Phase and Anomalous Hall Effect in a
frustrated ferromagnet. {\it Science} {\bf 291,} 2573 (2001).

[33] R. F. Wang {\it et al} Artificial `spin ice' in a
geometrically frustrated lattice of nanoscale ferromagnetic
islands. {\it Nature} {\bf 439,} 303 (2006).

[34] Castro Neto, A. H., Pujol, P. \& Fradkin, E. Ice: a strongly
correlated proton system. {\it Phys. Rev. B} {\bf 74,} 024302
(2006).

[35] Limelette, P. {\it et al.} Universality and Critical
Behaviour at the Mott Transition. {\it Science} {\bf 302,} 89
(2003).

[36] Kutnjak, Z., Petzelt, J. \& Blinc, R., The giant
electromechanical response in ferroelectric relaxors as a critical
phenomenon. {\it Nature} {\bf 441,} 956 (2006).

[37] Isakov, S. V., Wessel, S., Melko, R. G., Sengupta, K. \& Kim,
Y. B. Valence Bond Solids and Their Quantum Melting in Hard Core
Bosons on the Kagome Lattice. cond-mat/0602430 (2006).

\end{document}